\documentclass[prd,aps,
tightenlines,
superscriptaddress,showpacs,nofootinbib,eqsecnum,amsfonts,amsmath
]{revtex4}
\usepackage{bm,graphics,graphicx,epsfig,color
}
\def\be{\begin{equation}}
\def\ee{\end{equation}}

\def\gsim{\mathrel{
\rlap{\raise 0.511ex \hbox{$>$}}{\lower 0.511ex
\hbox{$\sim$}}}}
\def\lsim{\mathrel{
\rlap{\raise 0.511ex \hbox{$<$}}{\lower 0.511ex
\hbox{$\sim$}}}}

\begin{document}
\title{ 
Higher signal harmonics, LISA's angular resolution, and dark energy
}

\author{K.G.\ Arun} \email{arun@lal.in2p3.fr} 
\affiliation{LAL, Universit\'e Paris-Sud, IN2P3/CNRS, Orsay, France} \affiliation{${\mathcal{G}}{\mathbb{R}} \varepsilon{\mathbb{C}}{\mathcal{O}}$, Institut d'Astrophysique de Paris - C.N.R.S., Paris, France} 
\author {Bala R Iyer}\email{bri@rri.res.in}
\affiliation{Raman Research Institute, Bangalore, 560 080, India}
\author{B S Sathyaprakash}
\email{B.Sathyaprakash@astro.cf.ac.uk}
\affiliation{School of Physics and Astronomy, Cardiff University, 
5, The Parade, Cardiff, UK, CF24 3YB}
\author {Siddhartha Sinha$^{3,}$}\email {p_siddhartha@rri.res.in}
\affiliation{Dept. of Physics, Indian Institute of Science, Bangalore, 560 012, India.}
\author{Chris Van Den Broeck$^{4}$}\email{Chris.van-den-Broeck@astro.cf.ac.uk}
\noaffiliation{}

\begin{abstract}

It is generally believed that the angular resolution of the Laser 
Interferometer Space Antenna (LISA) for binary supermassive black holes 
(SMBH) will not be good enough to identify the host galaxy or galaxy cluster. 
This conclusion, based on using only the dominant harmonic of the 
binary SMBH signal, changes substantially when higher signal harmonics are 
included in assessing the parameter estimation problem. 
We show that in a subset of the source parameter space the angular 
resolution increases by more than a factor of 10, thereby making it 
possible for LISA to identify the host galaxy/galaxy cluster. 
Thus, LISA's observation of certain binary SMBH coalescence events could 
constrain the dark energy equation of state to within a few percent, 
comparable to the level expected from other dark energy missions. 

\end{abstract}

\date{\today}
\maketitle

\section{Introduction}

An outstanding issue in present day cosmology
is the physical origin of dark energy (see, e.g., Ref.\ \cite{PeeblesRatra03} 
for a review). Probing the equation-of-state-ratio ($w(z)$) 
provides an important clue to the question of whether
dark energy is truly a cosmological constant (i.e., $w=-1$).
Assuming the Universe to be spatially flat, a combination
of the Wilkinson Microwave Anisotropy Probe (WMAP) and Supernova Legacy Survey (SNLS) data yields significant
constraints on $w=-0.967^{+0.073}_{-0.072}$ \cite{Spergel07implication}.
Without including the spatial flatness as a prior into the analysis, 
WMAP, large-scale structure and supernova data place a stringent constraint 
on the dark energy equation of state, $w=-1.08\pm0.12$. The Laser 
Interferometer Space Antenna (LISA) could play an important role in 
investigating the nature of dark energy as argued in 
Refs.~\cite{HolzHugh05,DaHHJ06}.

Binary supermassive black holes (SMBH), often referred to as 
gravitational-wave (GW) ``standard sirens" (analogous to the 
electromagnetic ``standard candles") \cite{Schutz86}, are 
potential sources for the planned LISA mission. LISA would be 
able to measure the ``redshifted'' masses of the component
black holes and the luminosity distance to the source with good
accuracy for sources up to redshifts of a few.
However, GW observations alone cannot provide any information about
the redshift of the source. If the host galaxy or galaxy cluster is
known one can disentangle the redshift from the masses by optical 
measurement of the redshift. This would not only allow one to 
extract the ``physical" masses, but also provide an exciting 
possibility to study the luminosity distance-redshift relation 
providing a totally independent confirmation of the cosmological parameters. 
Further, this combined observation can be used to map the distribution 
of black hole masses as a function of redshift~\cite{Hughes02,BBW05a,BBW05b}.
For this to be possible, LISA should 
(a)~measure the luminosity distance to the source with a good accuracy
and (b)~localize the coalescence event on the sky with good angular 
resolution so that the host galaxy/galaxy cluster can be uniquely identified.

Refs.\ \cite{HolzHugh05,DaHHJ06} identified two potential 
problems in using binary SMBH as standard sirens. Firstly, they found 
that LISA's angular resolution might not be good enough to identify
the source galaxy or galaxy cluster, and that other forms of identification
would be needed, and secondly, they pointed out that weak lensing effects
would corrupt the distance estimation to the same level as LISA's systematic
error on the measurement of the luminosity distance. Their analyses, like
most other in the literature on LISA parameter estimation,  
were based on the  so-called {\it restricted} post-Newtonian (PN) waveforms.
The restricted waveforms (RWF) retain only the leading order (i.e., 
Newtonian) term in the wave amplitude, a PN series,  but incorporate 
the phase up to the maximum available PN order, which is currently 
3.5PN~\cite{BDIWW95,B96,BFIJ02,BDEI04}. Recent studies have shown that 
the inclusion of higher order amplitude terms in the waveform (and hence 
higher harmonics of the orbital frequency) would play an important role 
in the detection rates (by increasing the mass reach of the detector)
\cite{Chris06,ChrisAnand06,AISS07}
as well as in the problem of parameter estimation 
\cite{SinVecc00a,SinVecc00b,MH02,HM03,ChrisAnand06b} 
of both ground-based and space-based detectors. Specifically, 
Refs.~\cite{MH02,HM03} examined the improved angular resolution of different
space-based detector configurations due to the inclusion of higher harmonics.

In the present work, we revisit the problem of parameter estimation 
in the context of LISA using amplitude-corrected PN waveforms. 
We investigate systematically the variation in parameter estimation
with PN orders by critically examining the role of higher harmonics 
in the fast GW phasing, higher PN corrections in the amplitudes
and frequency sweep and their interplay with the slow modulations
induced due to LISA's motion. More importantly, we explore 
the improvement in the estimation of the luminosity 
distance and the angular parameters due to the inclusion of  higher 
harmonics in the waveform. We translate the error in the angular 
resolution to obtain the number of galaxies (or galaxy clusters) 
within the error box on the sky. We find that independent of the angular 
position of the source on the sky, higher harmonics improve LISA's 
performance on both counts raised in Refs.~\cite{HolzHugh05,DaHHJ06}: On the
one hand we will show that the angular resolution improves typically by
a factor of $\sim 2$-500 (greater at higher masses) and the error on 
the estimation of the luminosity distance goes down by a factor 
of $\sim 2$-$100$ (again, larger at higher masses). For many possible sky positions and orientations of the source, the inaccuracy in our
measurement of the dark energy would be at the level of a few percent, so that it would only be limited by weak 
lensing. We conclude that LISA could provide interesting constraints
on cosmological parameters, especially the dark energy equation-of-state,
and yet circumvent all the lower rungs of the cosmic distance ladder.

This paper is structured as follows. In the next section we introduce
our signal model and the LISA noise power spectral density we will use. 
In Section III we discuss our results on parameter estimation and their 
relevance for astrophysics and cosmology. Section IV gives an overview 
of various effects that are likely to affect
our estimates. Conclusions are presented in Section V. Technical details
on how the parameter estimation was performed can be found in Appendix A.
Finally, in Appendix B we give an in-depth discussion of the way parameter 
estimation is influenced by the inclusion of higher harmonics and their
amplitude corrections.

\section{Signal model and LISA noise power spectral density}

The post-Newtonian formalism has been used to study the evolution of 
a binary under gravitational radiation reaction 
to a very high order in the small parameter $v$ characterizing 
the velocity of the component objects, yielding accurate expressions 
for the orbital phase and the two gravitational wave polarizations.
For binaries consisting of component stars of negligible spin 
on quasicircular orbits, the most accurate computations currently 
known have corrections not only to the orbital phase  up to order $v^7$ 
(i.e., 3.5PN order in the notation of PN theory)~\cite{BDEI04,BDIWW95,BFIJ02,DIS01,DIS02}, 
but also corrections to the gravitational wave polarizations to order 
$v^5$ (i.e., 2.5PN order)~\cite{BIWW96,ABIQ04,KBI07}. 
We shall call this the ``full'' waveform (FWF).

The waveform as seen in LISA is modulated in two ways due to LISA's motion.
LISA consists of three spacecraft at the vertices of an equilateral 
triangle of 5 million kilometers, each craft on a heliocentric orbit
slightly inclined to the ecliptic. 
As the craft orbit the Sun, the triangular formation also spins 
around itself with the same one-year period as the orbital period. 
Therefore, relative to LISA the source location and orientation changes with time
with a one-year period and induces amplitude and phase 
modulations in the waveform. 

It is well-known that at signal frequencies $f \lesssim 5 \times 10^{-3}$ Hz, LISA can essentially be modeled as a pair of two-arm interferometers, usually labeled as I and II \cite{Cutler98}, and this suffices
for the sources considered in this paper. (However, it would be interesting to investigate the added value of the remaining third combination ignored in this
work.) In what follows, to begin with we consider a single detector.  

Let us consider a source of total mass $M = m_1 + m_2$ and symmetric mass ratio 
$\nu = m_1 m_2/M^2$ (where $m_1$, $m_2$ are the individual component masses) located at a luminosity distance $D_{\rm L}.$ 
In the stationary phase approximation (SPA), the Fourier transform 
$\tilde{h}_{\rm I}(f)$ of the response of detector I to the full waveform, including the modulations due
to LISA's motion, is given  
by~\cite{AISS07}:
\begin {equation}
\tilde{h}_{\rm I}(f) = \frac{\sqrt{3}}{2} \frac{2M\nu}{D_L}
\,\sum_{k=1}^{7}\,\sum_{n=0}^5\,
\frac{A^{\rm I}_{(k,n/2)}(t(f_k))
\,x^{\frac{n}{2}+1}(t(f_k))\,e^{-i\phi^{\rm I}_{(k,n/2)}(t(f_k))}}{2\sqrt{k\dot{F}(t(f_k))}}\,
\exp\left[i\,\psi_{f,k}(t(f_k))\right],
\label {FT}
\end {equation}
where $f_k\equiv f/k,$ an overdot denotes derivative with respect to time, and
$\psi_{f,k}(t(f_k))$ is given by
\begin {equation}
\psi_{f,k}(t(f_k)) = 2\pi f\,t(f_k) -
k\,\Psi(t(f_k))-k\,\phi_{\rm D}(t(f_k)) - \pi/4.
\label {phase}
\end {equation}
In Eq.\ (\ref{FT}), the factor $\sqrt{3}/2$ is due to the $60${\mbox{$^{\circ}$}}
angle between the interferometer's arms and $t(f)$ is given in Ref.\ \cite{BFIJ02}.
The waveform is a superposition of harmonics of the orbital frequency (labeled by the index $k$), and each 
harmonic has PN contributions to the amplitude (labeled by $n$; note that we can only go up to $n=5$, as no amplitude corrections are explicitly known beyond 2.5PN). As the PN order in amplitude is increased, more and more harmonics appear; at 2.5PN order there are seven, which is why the index $k$ only runs up to $k=7$. 
Quantities in Eqs.\ (\ref{FT}) and (\ref{phase}) with the argument $t(f_k)$
denote their values at the time when the instantaneous orbital frequency sweeps
past the value $f/k$ and $x(t)$ is the PN parameter given by
$x(t) = (2\pi M F(t))^{2/3}$, $F(t)$ being the instantaneous orbital frequency
of the binary. $A^{\rm I}_{(k,n/2)}(t)$ and $\phi^{\rm I}_{(k,n/2)}(t)$ are the polarization
amplitudes and phases of the $k$th harmonic appearing at the $n/2$th PN order.
$\Psi(t)$ is the orbital phase of the binary and $\phi_{\rm D}(t)$ is a time-dependent 
term representing Doppler modulation. Explicit expressions for $A^{\rm I}_{(k,n/2)}$ and $\phi^{\rm I}_{(k,n/2)}$ can be found in \cite{ChrisAnand06}; time-dependence of these quantities arises through the beam-pattern functions due to the varying sky position and orientation of the source relative to the detector \cite{Cutler98}. The expression for $\phi_{\rm D}(t)$ is given in \cite{Cutler98,AISS07}. For the PN expansions for $t(F)$, $\Psi(F)$, $\dot{F}(F)$  
we refer to Ref.\ \cite{BFIJ02}. 

The restricted waveform (RWF) corresponds to retaining the term with 
$k=2$ and $n=0$ in Eq.~(\ref{FT}) and neglecting all others. It is 
clear that the RWF has only the dominant harmonic at twice the orbital 
frequency but no other harmonic, nor PN corrections to the dominant
one. It does, however, include the post-Newtonian expansion of the phase
to all known orders, i.e., up to $v^7.$ The FWF, on the other hand,  has 
not only the dominant harmonic but also other harmonics up to seven 
times the orbital frequency and their PN corrections to the relevant 
order. The distinctive nature of the FWF as compared to RWF, especially 
the richer structure in its spectrum, can be seen in Fig.\ 1 
of Refs.\ \cite{ChrisAnand06b,AISS07}.

Expanding the denominator and extracting the lowest order term helps
us rewrite the waveform in a form more suitable for our purposes,
\begin {eqnarray}
\tilde{h}_{\rm I}(f) & = & \frac{\sqrt{5}}{8}\frac{1}{\pi^{2/3}}\frac{{\cal M}^{5/6}}{D_{\rm L}}
\frac{1}{(2f)^{7/6}}\sum_{k=1}^{7}k^{2/3}e^{i\,\psi_{f,k}(t(f_k))}\nonumber\\
&&\left[\left(\sum_{n=0}^5 A^{\rm I}_{(k,n/2)}(t(f_k))\,
(2\pi Mf_k)^{n/3}\,e^{-i\phi^{\rm I}_{(k,n/2)}(t(f_k))}\right)
\left(\sum_{m=0}^5 S_{(m/2)}\,(2\pi Mf_k)^{m/3}\right)\right]_p,
\label {FWF}
\end {eqnarray}
where ${\cal M} = M\, \nu^{3/5}$ is the chirp mass of the binary, 
and $[\,\cdot\,]_p$ denotes consistent truncation to $p$th post-Newtonian order 
(in our case $p=2.5$).  The coefficients $S_{(m/2)}$ are the PN expansion
coefficients of $\dot{F}(t(f_k))^{-1/2}$ and are given in Eq.~(A.4) of \cite{ChrisAnand06}.

Each harmonic in $\tilde h_{\rm I}(f)$ is taken to be zero outside a certain frequency range. The upper cutoff frequencies are dictated by the last stable orbit (LSO), beyond which the PN approximation breaks down. For simplicity we assume that this occurs when the orbital frequency $F(t)$ reaches $F_{\rm LSO} = 1/(6^{3/2} 2\pi M)$ -- the orbital frequency at LSO of a test particle in Schwarzschild
geometry in $c=G=1$ units\footnote{Note that the cutoffs are placed on the orbital
frequency of the binary, not the dominant GW harmonic; hence the extra factor of
2 in the denominator of the expression for $F_{\rm LSO}$.}. Consequently, in the frequency domain, the contribution to $\tilde{h}(f)$ from the $k$th harmonic is set to zero for frequencies above $k F_{\rm LSO}$. In determining the lower cutoff frequencies we assume that the source is observed for at most one year, and the $k$th harmonic is truncated below a frequency $k F_{\rm in}$, where $F_{\rm in}$ is the value of the orbital frequency one year before LSO is reached \cite{AISS07}:
\begin{equation}
F_{\rm in} = F(t_{\rm LSO} - \Delta t_{\rm obs}) = 
\frac{F_{\rm LSO}}{\left(1 + \frac{256\nu}{5 M}\Delta t_{\rm obs} v_{\rm LSO}^8\right)^{3/8}}.
\end{equation}
For simplicity the quadrupole formula was used. In the above, $t_{\rm LSO}$ and $v_{\rm LSO} = 1/\sqrt{6}$ are, respectively, the time and orbital velocity at last stable orbit, and $\Delta t_{\rm obs} = 1$ yr.
However, LISA's sensitivity becomes poorer and poorer below $\sim$3 mHz and current estimates normally assume a ``noise wall" at $f_{\rm s} = 10^{-4}$ Hz. Thus, we take the lower cutoff frequency of the $k$th harmonic to be the maximum of $f_{\rm s}$ and $k F_{\rm in}$. For more details we refer to \cite{AISS07}.   
 
As we mentioned before, the LISA detector can be viewed as a combination of two independent detectors. Until now we have dealt with one detector. Calculations for the two detector case closely follow the corresponding treatment for the RWF, which is explained in detail in \cite{Cutler98}. The beam-pattern functions for detector II can be obtained from those of detector I by a simple rotation. The waveform $\tilde{h}_{\rm II}(f)$ for the second detector is formally identical to $\tilde{h}_{\rm I}(f)$, and quantities $A^{\rm II}_{(k,n/2)}$ and $\phi^{\rm II}_{(k,n/2)}$ are obtained from their counterparts $A^{\rm I}_{(k,n/2)}$ and $\phi^{\rm I}_{(k,n/2)}$ by replacing the beam-pattern functions of detector I by those of detector II.  

The waveform given in Eq.\ (\ref{FWF}) together with its counterpart for the second detector form the basis of the analysis in this
paper. Following earlier works of 
Refs.~\cite{Cutler98,Hughes02,BBW05a,ALISA06,BarackCutler04}
we employ the Fisher matrix approach \cite{Finn92,FinnCh93} to the problem of parameter
estimation. The waveforms depend on nine parameters which are chosen to be
\begin{equation}
{\mathbf p} \equiv \left(\ln{\cal M},\delta,t_{\rm C},\phi_{\rm C},\ln D_{\rm L},
\mu_{\rm S},\mu_{\rm L},\phi_{\rm S},\phi_{\rm L}\right),
\label{eq:params}
\end{equation}
where $\delta\equiv (m_2-m_1)/M$ ($m_1$ and $m_2$ being the individual masses; throughout this 
paper we assume $m_2 \geq m_1$)\footnote{The waveform is invariant under interchange of mass labels \emph{provided} that at the same time, the phasing is shifted by $\pi$; since we use a single phasing formula we need an ordering in the definition of $\delta$. The parameter $\delta$ was originally introduced in place of $\nu$ to ensure that the FWF Fisher matrix remains nonsingular on the surface $m_1=m_2$. The situation is reversed in the case of RWF: there the Fisher matrix becomes singular in the limit $\delta \rightarrow 0$ whereas it remains regular at $m_1 = m_2$ when $\nu$ is used in place of $\delta$ (see the discussion in \cite{ChrisAnand06b}). The equal mass case is dealt with in more detail in \cite{SintesAndTrias}.} is
used as a parameter instead of $\nu$ following Ref.~\cite{ChrisAnand06b};
$t_{\rm C},\,\phi_{\rm C}$ are, respectively, the time and orbital phase at coalescence\footnote{Below we will consistently set the \emph{values} of $t_{\rm C}$ and $\phi_{\rm C}$ to zero, but both parameters are included as coordinates on the space of signals in computing the Fisher matrix.}; 
$\mu_{\rm S} = \cos \theta_{\rm S}$ and $\phi_{\rm S}$ determine the source position in the 
sky; while $\mu_{\rm L} = \cos \theta_{\rm L}$ and $\phi_{\rm L}$ determine the orientation
of the binary's orbit with respect to a nonrotating detector at the solar system
barycenter\footnote {This is a different notation from Ref.~\cite{Cutler98},
where the source angles measured in the fixed barycenter frame are denoted by
($\bar \theta_{\rm S},\, \bar \phi_{\rm S},\, \bar \theta_{\rm L},\, \bar \phi_{\rm L} $).}. Following Ref.~\cite{Cutler98}, we have fixed the initial position and
orientation of LISA by setting the constants $\phi_0$ and $\alpha_0$ defined there to zero at $t=0$.
The polarization amplitudes and phases depend on the location and orientation
of the source through the beam-pattern functions. The orientation of the source changes relative to LISA with the period of a year. 
Therefore, these quantities are modulated on a one-year time scale and
depend on $\mu_{\rm S}, \mu_{\rm L}, \phi_{\rm S},\phi_{\rm L}$  and also on $\delta$, $\ln {\cal M}$ and $t_{\rm C}$. 
The phase $\psi_{f,k}(t(f_k))$ depends on $\ln {\cal M}, \delta, t_{\rm C}$ and $\phi_{\rm C}$ and 
varies with the orbital timescale which is much shorter than a year. 

Following Ref.~\cite{HMD07}, the parameters can be subdivided into two subcategories:
${\mathbf p}_{\rm fast}=\left(\ln{\cal M},\delta,t_{\rm C},\phi_{\rm C}\right)$ and
${\mathbf p}_{\rm slow}=\left(\ln D_{\rm L},\mu_{\rm S},\mu_{\rm L},\phi_{\rm S},\phi_{\rm L}\right)$.
The classification arises naturally because the signal that LISA observes
can  be viewed as a (slow) low frequency modulation due to its motion 
around the Sun superposed on the (fast)  high (GW) frequency carrier signal.
The accuracy  of estimation of  ${\mathbf p}_{\rm fast}$ follows from the GW 
phasing of the binary while that of ${\mathbf p}_{\rm slow}$ from the  modulations 
associated with LISA's orbital motion.

In our analysis, we take the noise power spectral density (PSD) to
be that given by Eqns.~(2.28)--(2.32) of Ref.~\cite{BBW05a}. As usual, the Fisher matrix ${\mathbf \Gamma}$ for LISA as a whole is simply ${\mathbf \Gamma} = {\mathbf \Gamma}_{\rm I} + {\mathbf \Gamma}_{\rm II}$, where ${\mathbf \Gamma}_{\rm I, II}$ are the Fisher matrices computed from the waveforms $\tilde{h}_{\rm I, II}(f)$. The parameters used will be the ones listed in Eq.~(\ref{eq:params}), so that ${\mathbf \Gamma}$ is a $9\,\times\,9$ matrix. However, the errors in the 
estimation of $\mu_{\rm S}$ and $\phi_{\rm S}$ obtained in this way will be converted to 
a solid angle $\Delta \Omega_{\rm S}$ centered around the actual source direction.
Following the notation of \cite{BarackCutler04},
\begin{equation}
\Delta \Omega_{\rm S}=  2\pi \sqrt{(\Delta \mu_{\rm S} \,\Delta
\phi_{\rm S})^2-\langle\delta \mu_{\rm S}\,\delta \phi_{\rm S}\rangle^2},
\end{equation}
where the second term is the covariance between $\mu_{\rm S}$ and  
$\phi_{\rm S}$. A similar quantity $\Delta \Omega_{\rm L}$ is used to quantify the error estimate in the orientation of the binary's orbit.

In what follows, whenever it is necessary to consider a specific cosmological model we will assume a flat Universe with Hubble constant $H_0 = 75\,\mbox{km}\,\mbox{s}^{-1}\mbox{Mpc}^{-1}$, matter density $\Omega_{\rm m}=0.27$, and dark energy density $\Omega_{\rm d}=0.73$, with $\Omega_{\rm Total}=\Omega_{\rm m} + \Omega_{\rm d} = 1$.

\section{The results and their astrophysical implications}

In this Section we will discuss the extent to which LISA will be able to 
constrain cosmological parameters by observing a binary SMBH with a 
large signal-to-noise ratio and measuring their parameters, most 
importantly their location on the sky and the luminosity distance. As we shall see,
the number of clusters in LISA's angular error box reduces dramatically
when using the FWF as compared to RWF, thereby enabling us to identify
the host galaxy, or galaxy cluster. Consequently, it should be possible
to measure the dark energy equation of state by combining LISA observations
with electromagnetic observations.

We start by outlining the generic features of parameter 
estimation with higher harmonics; after that we will focus on
angular resolution and the estimation of luminosity distance, and what these
can tell us about cosmology.

\subsection{Parameter estimation with the full waveform: General trends}

Inclusion of higher harmonics results in a significant improvement in the 
determination of a binary's parameters in the context of LISA, as is the 
case for ground-based detectors \cite {ChrisAnand06b}.
A typical variation of parameter estimation (PE) accuracy with PN orders in amplitude
is displayed\footnote{The numerical values in all our tables
and results are unaffected by the missing terms discussed in 
\cite{KBI07} to the accuracies quoted.} in Table \ref{tab:PE1}.
We observe the following general features of amplitude-corrected 
waveforms with  regard to PE: For all masses 
and all angles we have explored, there is a significant improvement  
in the estimation of {\em all} parameters for LISA
 when considering the full waveform as compared to the restricted PN
waveform. 
\begin {table}[ht]
\centering
\begin{tabular}{|l|c|c|c|c|c|c|c|c|}
\hline
\hline
 PN  & \multicolumn {8}{c|}{$(10^6,\, 10^7) M_\odot$; {\bf $z=0.55$ }; 
 $\mu_{\rm S}=-0.8,\phi_{\rm S}=1,\mu_{\rm L}=0.5,\phi_{\rm L}=3$.}\\
\cline{2-9}
order& \multicolumn{1}{c|}{SNR}  & \multicolumn {1}{c|}{~$\Delta\,\ln{\cal M}$~} & \multicolumn {1}{c|}{~~~$\Delta\,\delta$~~~} & \multicolumn {1}{c|}{~~~$\Delta\,t_{\rm C}$~~} & \multicolumn {1}{c|}{~~$\Delta\,\phi_{\rm C}$~~~}  & \multicolumn{1}{c|}{~$\Delta \ln D_{\rm L} $~}& \multicolumn {1}{c|}{$\Delta\,\Omega_{\rm S}$} & \multicolumn {1}{c|}{$\Delta\,\Omega_{\rm L}$}\\
 & & $(10^{-6})$ &  $(10^{-6})$ & (sec) &(rad) & ($10^{-3}$) & ($10^{-5}$str) & ($10^{-5}$str)\\
\hline
0   & 1824 & 380& 310 & 90 & 65 & 32 & 2400  & 6600 \\
0.5 & 2005 & 110 & 110 & 32 & 2.6 & 2.3 & 6.3 & 9.9 \\
1   & 1793 & 87 &  93 & 28 & 2.6 & 2.5 & 6.1 & 8.0  \\
1.5 & 1680 & 87 &  95 & 29 & 2.9 & 2.7 & 6.8 & 8.7  \\
2   & 1585 & 94 & 100 & 31 & 3.1 & 2.8 & 7.8 & 10   \\
2.5 & 1549 & 96 & 100 & 31 & 3.2 & 2.9 & 8.2 & 11   \\
\hline
\hline
\end{tabular}
\caption{Variation of parameter estimation errors with post-Newtonian orders in amplitude
for a $(10^6, 10^7) M_\odot$ binary at  $z=0.55$ (corresponding to a luminosity distance of $D_L=3$ Gpc for 
a Hubble constant $H_0=75\,\mbox{km}\,\mbox{s}^{-1}\mbox{Mpc}$, matter density $\Omega_{\rm m}=0.27$, and dark energy density $\Omega_{\rm d}=0.73$, with $\Omega_{\rm Total}=\Omega_{\rm m} + \Omega_{\rm d} = 1$). The angles are chosen to be $\mu_S=-0.8$, $\phi_S=1$, $\mu_L=0.5$, $\phi_L=3$.}
\label{tab:PE1}
\end {table}
The orbital frequency and the inspiral rate, and therefore the phase
evolution of the waveform, are determined primarily by the chirp mass
${\cal M}$. Thus accurate phase tracking leads to a precise measurement 
of ${\cal M}$. The phase also depends on the 
``mass difference" $\delta$ which, therefore, can also be measured quite
accurately. This is borne out by Table~\ref{tab:PE1} 
where the trend is shown to be true for the FWF also. Finally, note the spectacular
improvement in angular resolution and the determination of luminosity distance in
going from RWF to FWF, which will be the focus of the rest of this Section. 

We relegate to Appendix B a more critical and in-depth discussion of the trends observed 
with increasing PN order in amplitude and consequent inclusion of higher harmonics in the waveform.

\subsection{Effect of higher harmonics on angular resolution and luminosity distance}

Table \ref{tab:PE8} lists the one-sigma errors in parameters of interest
for seven different combinations of the angular parameters, as in Ref.~\cite{Cutler98}, 
each for three different binary masses. The angular parameters (cf. A1, 
\ldots, A7 in the Table) chosen are a coarse sample of the possible orientations of
the orbit and the source's sky location and our choice of masses is indicative of 
the different binary SMBH coalescences LISA is likely to observe with a large 
signal-to-noise ratio (SNR). To make direct contact with astrophysical systems,
we give the \emph{physical} masses $M=M_{\rm phys}$ and not the observed (i.e.,
redshifted) masses $M_{\rm obs}$. The two are related by $M_{\rm obs} = 
(1+z) M_{\rm phys},$ where $z$ is the cosmological redshift of the source. In
order to compute the upper frequency cutoff one should first convert the total
physical mass to total observed mass and then use the formula for the LSO frequency.
Our sources are all at $z\simeq 0.55$, i.e., a luminosity distance of $D_{\rm L}
= 3\, \rm Gpc$. Theoretical predictions of event rates for SMBH coalescence vary over a
wide range, but the rate could be as high as 1 per year within $z=0.55$ \cite{Nelemans06}. (See
Section IV for a discussion of our results for merger events occurring at a higher redshift.) 
The physical masses and the corresponding 
LSO frequencies in the form $(m_1/M_\odot,\, m_2/M_\odot,\, 7F_{\rm LSO}/\rm mHz),$
are  $(10^5,\, 10^6,\, 9.03),$ $(6.45 \times 10^4,\, 1.29 \times 10^6,\, 7.33)$ and 
$(10^6,\, 10^7,\, 0.903)$. Thus, the highest harmonic (at seven times the orbital frequency) of the heaviest system that we consider does not quite reach the sweet spot of LISA's sensitivity. For the other two systems the dominant harmonic is close to the sweet spot and higher harmonics sweep through LISA's sensitivity bandwidth.

The table lists the 1-sigma errors incurred in the estimation of all the 
parameters except for the errors on $\mu_{\rm L},$ $\phi_{\rm L}$ (the direction of orbital 
angular momentum) and $\phi_{\rm C}.$ As usual, we have converted the error 
in the estimation of $\mu_{\rm S}$ and $\phi_{\rm S}$ to an error in the 
solid angle $\Delta \Omega_{\rm S}$ centered around the actual source direction. 
For the sake of completeness we have given, in alternate rows, the errors 
for both the RWF and FWF.

For certain values of the angular parameters the presence of the harmonics seems to have a 
considerable impact on the determination of the luminosity distance and the 
angular position of the source.  The errors in the luminosity distance
(i.e., $\Delta D_{\rm L}$) and the source's sky position 
(i.e., $\Delta \Omega_{\rm S}$) are reduced by factors up to 600 and
400, respectively, while using FWF as compared to RWF. 
This means that the error box could be smaller by a 
factor of $2.4\times 10^5$. However, what is relevant for cosmological 
applications is by how much the error in the sky position goes down 
(i.e., about a factor of 2.5 to 400).
Interestingly, the heavier, and astrophysically most relevant, system 
of $(10^6,10^7)\,M_\odot$, where none of the harmonics get close to the detector's 
sweet spot, shows the largest improvement in distance estimation and angular resolution in 
going from restricted to amplitude-corrected waveforms. This observation, based only 
on the few systems studied in this paper, is found to be generally true in 
an independent and more exhaustive study by Trias and Sintes \cite{SintesAndTrias}. The larger improvement over the RWF is to be expected since for very massive systems only the higher harmonics radiate significantly within the detector's bandwidth. However, what is striking is that when considering only FWF, for most choices of angles the errors on distance and the angular resolution are almost at the same level as for lighter systems. 
Doppler modulation does not seem to affect the accuracy of estimation of parameters
for the systems considered in this paper; they are expected to be important for 
systems with lower masses \cite{Cutler98}. An alternative method to study these issues 
is by a direct use of the three time delay interferometry variables. Work along these lines is in progress
\cite{BPS07} and should provide an independent check of our results in the near future.

\begin{table}
	\centering
		\begin{tabular}[t]{c|c|c|c|c|c|c|c|c|c|c|c|c|c}
		\hline
		\hline
     Orientation & $\mu_{\rm S}$ & $\varphi_{\rm S}$  & $\mu_{\rm L}$ & $\varphi_{\rm L}$ & Model &   SNR  & $\Delta \ln D_{\rm L}$ 
     & $\Delta \Omega_{\rm S}$ & $\Delta \ln {\cal M}$&  $\Delta \delta$  
     & $\Delta t_{\rm C}$ & $\mathbf{N_{\rm clusters}}$ & $\Delta w$ \\ 
     &  & rad & & rad   &   &  & ($10^{-2}$)   &   ($10^{-6} \rm str$)   & ($10^{-6}$)      
     & ($10^{-6}$) & sec &  & \\     
     \hline
     \multicolumn{14}{l}{$(m_1,\, m_2) = (10^5,\, 10^6)M_\odot $}\\
     \hline
     A1            & 0.3    & 5 & 0.8   & 2 &  RWF & 750  & 1.2  & 12   & 6.0 & 31  & 1.7   & \bf 0.25 & 0.068 \\
                   &        &   &       &   &  FWF & 754  & 0.88 & 4.3  & 4.6 & 23  & 1.2   & \bf 0.088 & 0.050 \\
\hline
     A2            & $-0.1$ & 2 & $-0.2$ & 4 & RWF &1168 & 1.1  & 110   & 4.7 & 21 & 1.7  & \bf 2.2 &  0.062  \\
                   &        &   &        &   & FWF &1150 & 0.58 & 13  & 3.5 & 16 & 1.1  & \bf 0.27 &   0.033  \\
\hline   
     A3            & $-0.8$ & 1 & 0.5   & 3 & RWF  &2722  & 0.25 & 170  & 3.3 & 12  & 2.6   & \bf 3.5 &  -- \\
                   &        &   &       &   & FWF  &2497  & 0.17 &  26  & 2.7 & 9.7 & 1.1   & \bf 0.53&  0.0096 \\   
\hline
     A4            & $-0.5$ & 3 & $-0.6$&$-2$& RWF  &1868  & 0.74 & 150   & 3.1 & 15 & 1.2   & \bf 3.1 & -- \\
                   &        &   &       &    & FWF  &1781  & 0.19 & 13  & 2.5 & 12 & 0.58  & \bf 0.27  & 0.011 \\
\hline
     A5            & 0.9    & 2 & $-0.8$& 5 & RWF  &3740  & 15 & 84   & 2.3 & 8.0 & 2.1   & \bf 1.7 & 0.82 \\
                   &        &   &       &   & FWF  &2857  & 0.11 & 8.1  & 1.7 & 7.9 & 0.69  & \bf 0.17 & 0.0062 \\        
\hline
     A6            & $-0.6$ & 1 & 0.2   & 3 & RWF  &2185  & 0.42 & 220  & 3.9 & 15  & 2.9   & \bf 4.5 &  -- \\
                   &        &   &       &   & FWF  &2108  & 0.24 & 65   & 3.0 & 11  & 1.6   & \bf 1.3 &  0.014 \\      
\hline
     A7            & $-0.1$ & 3 & $-0.9$& 6 & RWF  &2213  & 0.58 & 410  & 3.5 & 13  & 1.1   & \bf 8.4 &  -- \\
                   &        &   &       &   & FWF  &2175  & 0.45 & 300  & 2.9 & 10  & 0.74   & \bf 6.1 & -- \\  
     \hline       
     \multicolumn{14}{l}{$(m_1,\, m_2) = (6.45\times 10^4,\,1.29\times 10^6)M_\odot $}\\
     \hline
     A1            & 0.3    & 5 & 0.8   & 2 & RWF  & 385  & 1.3  & 21   & 5.5 & 13  & 3.2  & \bf 0.43 & 0.073 \\
                   &        &   &       &   & FWF  & 511  & 1.0  & 8.4  & 4.2 & 9.1 & 2.1  & \bf 0.17 & 0.056 \\
\hline
     A2            & $-0.1$ & 2 & $-0.2$& 4 & RWF  & 595  &  1.1 &  120 & 4.2 & 9.2 & 2.5  & \bf 2.4  & 0.062 \\
                   &        &   &       &   & FWF  & 771  & 0.70 &  25  & 3.3 & 6.5 & 1.7  & \bf 0.51 & 0.039 \\
\hline
     A3            & $-0.8$ & 1 & 0.5   & 3 & RWF  &1345  & 0.33 & 170  & 3.4 & 5.8 & 2.7  & \bf 3.5  & -- \\
                   &        &   &       &   & FWF  &1573  & 0.25 &  53  & 2.6 & 4.2 & 1.6  & \bf 1.1  & 0.014 \\
\hline
     A4            & $-0.5$ & 3 & $-0.6$&$-2$&RWF  & 924  & 0.78 & 160  & 3.0 & 6.8 & 1.7  & \bf 3.3  & -- \\
                   &        &   &       &   & FWF  &1158  & 0.26 & 27   & 2.3 & 5.0 & 1.0  & \bf 0.55 & 0.015 \\
\hline
     A5            & 0.9    & 2 & $-0.8$& 5 & RWF  &1863  & 15 &  87  & 2.4 & 3.8 & 2.2  & \bf 1.8  & 1.0  \\
                   &        &   &       &   & FWF  &1506  & 0.19 & 25   & 2.0 & 3.9 & 1.3  & \bf 0.51 & 0.011  \\
\hline
     A6            &  $-0.6$& 1 & 0.2   & 3 & RWF  &1069  & 0.47 & 240  & 4.1 & 7.2 & 3.1  & \bf 4.9  & --  \\
                   &        &   &       &   & FWF  &1378  & 0.32 & 110  & 2.9 & 4.8 & 2.1  & \bf 2.2  & 0.018 \\
\hline   
     A7            & $-0.1$ & 3 &$-0.9$ & 6 & RWF  &1093  & 0.57 & 420  & 3.1 & 6.1  & 1.6  & \bf 8.6 & -- \\
                   &        &   &       &   & FWF  &1448  & 0.50 & 350  & 2.5 & 4.2  & 1.1  & \bf 7.1 & -- \\                          \hline       
     \multicolumn{14}{l}{$(m_1,\, m_2) = (10^6,\, 10^7)M_\odot $}\\
     \hline
     A1            & 0.3    & 5 & 0.8   & 2 & RWF  &495   & 11 & 600  & 1400 & 1100 & 290 & \bf 12 & -- \\
                   &        &   &       &   & FWF  &444   & 2.2  & 16   & 190  & 240  & 75  & \bf 0.33 & 0.12 \\  
\hline
     A2            & $-0.1$ & 2 &$-0.2$ & 4 & RWF  &773  & 10  &  6500   & 870   & 710  & 190  & \bf 130 & -- \\
                   &        &   &       &   & FWF  &685  & 1.2   &  43     & 130   & 160  & 51   & \bf 0.88  & 0.068 \\
\hline
     A3            & $-0.8$ & 1 & 0.5   & 3 & RWF  &1824  & 3.2  & 24000& 380  & 310  & 90  & \bf 490  & -- \\
                   &        &   &       &   & FWF  &1549  & 0.29 & 82   & 96   & 100  & 31  & \bf 1.7  & 0.016 \\ 
\hline
     A4            & $-0.5$ & 3 &$-0.6$ &$-2$&RWF  &1249  & 6.9  & 2400 & 550  & 450  & 120 & \bf 49  & -- \\
                   &        &   &       &   & FWF  &1081  & 0.34 & 40   & 110  & 130  &  40 & \bf 0.82& 0.019 \\
\hline
     A5            & 0.9    & 2 & $-0.8$& 5 & RWF  &2493  & 110 & 8300 & 270  & 220  & 63  & \bf 170  & -- \\
                   &        &   &       &   & FWF  &1954  & 0.18 & 18   & 200  & 180  & 49  & \bf 0.37 & 0.010    \\ 
\hline
     A6            & $-0.6$ & 1 &  0.2  & 3 & RWF  &1465   & 4.7   & 53000   & 470   & 380  & 110  & \bf 1100   & -- \\
                   &        &   &       &   & FWF  &1273   & 0.44  & 300     & 105   & 115  & 36   & \bf  6.1 & -- \\
\hline
     A7            & $-0.1$ & 3 & $-0.9$& 6 & RWF  &1480  & 21 & 170000 & 520& 390  & 98  & \bf 3500 & -- \\
                   &        &   &       &   & FWF  &1300  & 1.3  &  3400  & 87 & 100  & 30  & \bf 69   &  -- \\ 
   \hline
     \hline
		\end{tabular}
	\caption{Accuracy in LISA's measurement of the various parameters in Eq.~(\protect{\ref{eq:params}}),
	for seven different sets of the angular parameters and three different combinations of the (physical) masses at a distance of 3 Gpc ($z=0.55$).
When the number of clusters in the error box on the sky is significantly larger than 1, it will not be possible to determine redshift unless the inspiral event has a clear optical counterpart; we have chosen not to quote results for $\Delta w$ in such cases. (Note that the error on $w$ is ultimately determined by both LISA's statistical errors and weak lensing errors in the determination of luminosity distance.)  
	The figures clearly demonstrate  significant improvement in parameter estimation when higher order
	terms are included.}
	\label{tab:PE8}
\end{table}

\subsection{Number of clusters in LISA's error box}

Binary black holes are standard sirens \cite{Schutz86}. The amplitude of
gravitational waves from a binary SMBH is proportional to ${\cal M}^{5/6}/D_{\rm L}.$  
As evidenced by the numbers in Table \ref{tab:PE8},
LISA will measure both the chirp mass $\cal M$ of the source 
and the amplitude of gravitational waves
to a great precision. Thus, the luminosity distance to a source 
can be extracted by gravitational-wave observations alone.
In order to derive the luminosity distance-redshift relation, it is also necessary
to measure the redshift $z$ to the source, but LISA cannot measure $z$. However,
it might be possible to determine the source's redshift if the host galaxy,
or galaxy cluster, can be optically identified. Whether or not this is possible 
depends on how good LISA's angular resolution is, and whether it is small enough that no 
more than a few galaxies or galaxy clusters are found within the angular 
error box\footnote{
Note that we have to use the \emph{angular} error box and cannot use the smaller \emph{volume} error 
box also fixed by the luminosity
distance. In order to do precision cosmology we have to measure the source's
redshift {\em independently} of the luminosity distance.}.

To determine the number of galaxy clusters within a solid angle $\Delta\Omega_{\rm S}$ 
of the angular error box we need the comoving volume corresponding to a cone
defined by $\Delta\Omega_{\rm S}$ whose height is the physical distance from the 
detector to the source, which, of course, depends on the cosmological model. In a 
Universe whose matter density is $\Omega_{\rm m}$ and in which dark energy density takes the 
form of a cosmological constant\footnote{In the next subsection we will relax this assumption.} $\Omega_{\rm d}$ 
(with $\Omega_{\rm Total} = \Omega_{\rm m} + \Omega_{\rm d}=1$), the comoving volume per 
unit redshift within a box of angular size $\Delta \Omega_{\rm S}$ is:
\begin{equation}
\frac{dV_{\rm C}(z)}{dz} = \frac{\Delta\Omega_{\rm S}}{H_0} 
\frac{D_{\rm L}^2(z)}{(1+z)^2}\frac{1}{\sqrt{\Omega_{\rm m}(1+z)^3 + \Omega_{\rm d}}} \;.
\label{eq:diffVolume}
\end{equation}
In the above, $D_{\rm L}(z)$,
 the luminosity distance as a function of redshift, is given by
\begin{equation}
D_{\rm L}(z) = \frac{1+z}{H_0} \int_0^z \frac{dz'}{\sqrt{\Omega_{\rm m}(1+z')^3 + \Omega_{\rm d}}}\;,
\label{DL}
\end{equation}
where $H_0$ is the Hubble parameter at the current epoch.
The comoving volume from the observer to the source within a cone defined by 
$\Delta\Omega_{\rm S}$ is simply the integral of Eq.\ (\ref{eq:diffVolume}):
\begin{equation}
V_{\rm C}(z) = \int_0^z dz'\, \frac{\Delta\Omega_{\rm S}}{H_0} 
\frac{D_{\rm L}^2(z')}{(1+z')^2}\frac{1}{\sqrt{\Omega_{\rm m}(1+z')^3 + \Omega_{\rm d}}}\;. 
\label{eq:comovingVolume}
\end{equation}

The number density of clusters at high redshifts is not known very well. Following 
Ref.\ \cite{Bahcall:2003} we assume that the number density of clusters is $\sim 2 \times 10^{-5}
h^3 {\rm Mpc}^{-3},$ where $h$ is the present value of the Hubble parameter in
units of $100\, {\rm km}\, {\rm s}^{-1} {\rm Mpc}^{-1}.$ We take $h=0.75,$
$\Omega_{\rm m}=0.27$ and $\Omega_{\rm d}=0.73.$ For this choice of cosmological 
parameters the number of expected clusters within the LISA angular error box is
given in the second last column of Table \ref{tab:PE8}. Clearly, in many cases the number of 
clusters is of order 1 which means that LISA will help us to uniquely identify the 
host galaxy cluster of a binary SMBH merger event within a redshift of $z=0.55.$

There is a caveat with regard to the number of clusters found within the error 
box that is important to mention at this stage. Note that we integrated the comoving
volume up to $z=0.55,$ the location of our source. In reality, we would not know the 
redshift of the source and are not fully justified in integrating only up to this point; 
in principle we should consider all clusters within the error box up to much larger redshifts. 
However, sources at redshifts much larger than $z=0.55$ for the same luminosity distance would probably give radically different cosmological parameters. Consistency with other observations justifies considering only galaxy clusters that are roughly in the redshift region determined by 
inverting a luminosity distance-redshift relation based on parameter values from other measurements. In any case, we have checked that integrating the comoving volume up to $z=1$ (which for our chosen value of $D_{\rm L}$ would already imply a very significant departure from current cosmological models) does not drastically change the results of Table \ref{tab:PE8}.
Indeed, despite the higher limit of integration, the number of clusters in the angular error box remains less than 3 for most choices of angles, which is the (arbitrary) cutoff in ${\mathbf N}_{\rm clusters}$ we have chosen to assess whether redshift can be determined.

There have been suggestions that in order to identify the source associated with a binary
SMBH merger we should also look for optical/UV counterparts; the improved angular resolution with the
correct signal model should help in this case too. Optical and other electromagnetic 
telescopes will need to survey a much smaller area on the sky than was thought before
and should therefore more easily identify the galaxy cluster in which the merger took
place.

The error in $D_{\rm L}$ being less than a percent means that we should, in principle,
be able to tightly constrain the cosmological model a lot better. However, as discussed
by several groups, the possible effect of weak gravitational lensing on parameter
estimation, in particular on our ability to measure the luminosity distance (see, e.g., Ref.\
\cite{HolzHugh05}), will limit the extent to which LISA can measure dark energy.
These considerations do not alter the main conclusions of this paper as our main goal is to show
that the field of view in LISA's angular resolution in most cases involves only a few sources.

\subsection{Constraints on the equation-of-state of dark energy}

We conclude by mentioning the implications of our estimates
for astrophysics and cosmology. Up to this point in our analysis
we have not assumed any electromagnetic counterpart to the
coalescence events LISA will observe. But in reality,
many of the LISA observations are likely to have 
electromagnetic counterparts, either as a precursor
or as an afterglow~\cite{MiloPhinney05,ArmitageNatarajan02}.
The implications of an uniquely identified quasar source
in coincidence with LISA were examined in \cite{KFHM06} specifically
assuming quasars as a possible electromagnetic counterpart. Even
a single event of this type would provide us with unprecedented
tests of SMBH accretion physics, such as a precise measurement of the
Eddington ratio~\cite{KFHM06}.
For this to be possible, and to identify an unique electromagnetic
event in coincidence with a merger event as seen by LISA, the use
of higher harmonics would be crucial since for many
of the directions in the sky it brings down the number
of candidate clusters to less than one.

Another exciting possibility is to use LISA
as a cosmological probe. If a unique host is identified
electromagnetically in coincidence with the LISA observation, the 
redshift of the host galaxy will be known to very high accuracy.
The improved estimate of the luminosity distance obtained
by using the FWF would play a crucial role in determining the
cosmological parameters as suggested by Schutz~\cite{Schutz86}.

Gravitational-wave observation of a single inspiral event coupled with an electromagnetic determination of the redshift would allow LISA to strongly constrain the equation of state of dark energy. In a flat Universe, the luminosity distance can be written as
\begin{equation}
D_{\rm L} = (1+z)\,\int_0^z \frac{dz'}{H(z')},
\label{DLgeneral}
\end{equation}
where $H(z)$ is the Hubble parameter.
Given a form of matter energy with 
density parameter $\Omega_{\rm d}$ and a (constant) equation-of-state index $w$, one has
\begin{equation}
H(z) = H_0 \left[ \Omega_{\rm m}(1+z)^3 + \Omega_{\rm d} (1+z)^{3(1+w)} \right]^{1/2}.
\label{Hz}
\end{equation}
For a given, fixed redshift (and fixed $H_0$, $\Omega_{\rm m}$, and $\Omega_{\rm d}$), the error on $w$ is
\begin{equation}
\Delta w  = D_{\rm L} \left|\frac{\partial D_{\rm L}}{\partial w} \right|^{-1} \frac{\Delta D_{\rm L}}{D_{\rm L}}.
\label{Deltaw}
\end{equation}
Using (\ref{DLgeneral}) and (\ref{Hz}) and setting $\Omega_{\rm d} = 0.73$, $w= -1$, and $D_{\rm L} = 3$ Gpc, 
one obtains
\begin{equation}
\left|\frac{\partial D_{\rm L}}{\partial w} \right| \simeq 533 \, \mbox{Mpc}.
\label{dDLdw}
\end{equation}
With these assumptions and using the values for $\Delta D_{\rm L}/D_{\rm L}$ from our analysis, 
we find that in all of our examples, the FWF always leads to a smaller value for $\Delta w$ than the RWF whenever a comparison can be made. Indeed,
in Table \ref{tab:PE8} there are many instances where the RWF doesn't allow for a determination of the redshift because the number of clusters in the
angular error box on the sky is too large, in which cases $w$ cannot be measured. In most of the examples we have considered, the FWF does not have this problem.

The foregoing analysis does not take into account the error in luminosity
distance arising as a result of weak lensing of gravitational waves by the
intervening mass concentrations between the binary SMBH source and LISA. It is
estimated that the weak lensing will introduce errors in the luminosity distance
at the level of about 3-5\% for sources at $z\sim 0.5$ \cite{HolzHugh05}. This is far greater than
systematic error in LISA's measurement of the luminosity distance. Therefore, weak
lensing will be the limiting factor in LISA's ability to measure the dark energy
equation of state unless weak lensing can be corrected by properly modeling the
weak lenses.

We now revisit the caveat in the previous subsection regarding the fact that we tend to consider an error cone 
that stretches only to $z=0.55$. Let us evaluate $\Delta w$ when considering a measured luminosity distance of 
$D_{\rm L} = 3$ Gpc, but allowing for the possibility that the source may be at a different redshift. Given our 
luminosity distance, to assume that the source could be at, e.g., $z=0.6$ and keeping $H_0$, $\Omega_{\rm m}$ and 
$\Omega_{\rm d}$ the same would already require $w=-0.47$ (from Eqs.~(\ref{DLgeneral}) and (\ref{Hz})), a value excluded 
by WMAP and SNLS results \cite{Spergel07implication}. In that case $|\partial D_{\rm L}/\partial w| \simeq 668$ Mpc; 
substituting this into the right hand side of Eq.~(\ref{Deltaw}) one finds that all values of $\Delta w$ will 
\emph{decrease} by about 25 percent compared with the $z=0.55$ ($w=-1$) case. In reality, uncertainties in 
$H_0$, $\Omega_{\rm m}$, and $\Omega_{\rm d}$ will also have to be taken into account, but it will probably not be 
necessary to consider potential sources in the angular error box at redshifts that differ from the ``favored" 
value by more than 20 percent. A more complete treatment should take into account the covariances among $H_0$, 
$\Omega_{\rm m}$, $\Omega_{\rm d}$, and $w$; this we relegate to a future study.

We conclude by noting another interesting feature of our analysis. In addition to the improved angular resolution and luminosity distance, the 
errors in the estimation of mass parameters reduce considerably while using FWF. 
Even though the RWF itself would give a very good estimate of the masses,
the improved measurement of the mass parameters would be very important in 
performing certain novel tests of general theory relativity and its 
alternatives~\cite{BBW05a,AIQS06a,AIQS06b}. For example, the use of FWF should
help improve the accuracy with which the individual phasing coefficients can be 
determined independent of each other, an idea proposed in Ref.~\cite{AIQS06a,AIQS06b}.

\section{Factors affecting the estimates}
In this Section we will briefly discuss some caveats regarding our results. Here we focus on 
physical issues; limitations related to our chosen methods for computing errors (and their resolution) are commented on in Appendix A. 

\paragraph {Effect of black hole spins.}
Our analysis is restricted to the case of nonspinning black holes
while astrophysical black holes are known to have significant amounts 
of spin. Including the spin effect could significantly
alter the estimation of different parameters~\cite{BBW05a}.
To get an estimate of the effect of the
spin on parameter estimation, we have recomputed the covariance matrix by including the leading order spin-orbit
parameter in the waveform model.
We observed a deterioration of up to an order of magnitude for
the mass parameters but only a factor of a few in the
estimation of luminosity distance and angular resolution.
As suggested by Vecchio~\cite{Vecchio04}, the inclusion of precession
should compensate for the deterioration of the mass errors to a great extent. For
instance, including the leading spin-orbit term with a simple precession model,
Ref.~\cite{Vecchio04} showed that the angular resolution and luminosity
distance determination could be improved by a factor between 3-10.
A more recent analysis~\cite{LangHughes06} incorporated the 2PN spin
effects. Thus our estimates may not be too far from the realistic case
despite the assumption of nonspinning holes. 
An interesting exercise for the future would be to include
the amplitude corrections with the spin effects and see the extent
to which the results of this paper change.

\paragraph {Weak and strong gravitational lensing.} 
An important physical effect we have neglected in our 
analysis is the possibility of gravitational lensing of the signal.
Weak lensing by intervening matter distributions distorts the
gravitational waveform, inducing a systematic error in the
estimation of luminosity distance ~\cite{HolzHugh05}.
This could be as high as 5-10\% for some of the systems, much
higher than LISA's systematic error on the luminosity distance.
On the other hand, strong lensing may
improve the angular resolution, as multiple gravitational-wave ``images"
 formed will reduce the correlations between various
 parameters~\cite{TakaNaka03}.
This improvement could be as high as 100 times for a million solar
mass (redshifted) binary~\cite{Seto04}. Thus the net effect of lensing
of a high redshift source may not be disastrous.

\paragraph {Inaccuracy in the cosmological parameters.}
In computing the accuracy with which we can measure the
dark energy parameter $w$ we have assumed that other
cosmological parameters, namely $H_0,$ $\Omega_{\rm m},$
and $\Omega_{\rm d},$ are all known precisely and that
the redshift $z$ of the host galaxy is measured accurately.
In reality, all these quantities will have their own
statistical errors that must be folded into the analysis
and we hope to do so in a future publication.  Moreover,
it might be possible for LISA to measure the luminosity
distance-redshift curve by observing many supermassive
black hole binary coalescences during its mission lifetime
and extract all the cosmological parameters. How well LISA
might be able to do would be a very interesting question
that we would like to reserve for future studies. In any
case, the fact that higher harmonics enable a greater
accuracy in the measurement of the parameters should help.

\paragraph {Event rates and measurement of $w$.} 
The rate of binary SMBH mergers in the Universe is not known with
good accuracy. The rate depends on the cosmological model of choice
and on structure formation and growth. Current models predict rates
that vary over two orders of magnitude. For one of the models the rate
of binary SMBH merger is as large as $\sim 10\,\mbox{yr}^{-1}$ within $z=1$ or about
$\sim 1\,\mbox{yr}^{-1}$ up to $z=0.55$ (see \cite{Nelemans06} for an overview). A single source
at the right point in parameter space at $z=0.55$ will be good 
enough to measure $w$ to a few percent. However, events at $z=1$ will
be more frequent and consequently a few of them are likely to end up 
in the favored region of parameter space.
We have verified that at $z=1$, for all choices of masses considered and angles
A1--A5 (except for A3 in the case of the most massive system), the number of clusters within the error box is less than 10 when using the FWF. It should therefore be possible to either directly identify
the merger source (by demanding consistency of the inferred cosmological parameters
with those from other observations), or to locate it by observing an afterglow. Values of $\Delta w$ will be about 70 percent higher than before; however, with repeated observations one would be able to perform a statistical analysis along the lines of \cite{Schutz86}. Thus, merger
events at $z=1$ also allow for a measurement of $w$, although in this case the
microlensing would probably restrict our accuracy in its estimation to 
$\sim 10$ percent \cite{HolzHugh05}.

\section{Conclusions}
In this paper we have reassessed LISA's ability to perform precision 
cosmology by using waveform models with far richer structure 
than those used before, and we find remarkable improvement in its angular
resolution and estimation of the luminosity distance. 
Our ability to reliably measure the parameters of a binary SMBH depends 
crucially on the signal model onto which the data is projected. If 
the signal model is incorrect or even inaccurate, then regardless 
of how good the instrument might be, the measurement is prone to 
systematic errors which will make precision measurements meaningless (see, e.g., Ref.~\cite{CutlerVallisneri07} for a detailed discussion on parameter extraction errors due to inaccurate
template waveforms).
From the point of view of incurring the minimum possible errors in the 
estimation of parameters it is desirable to use the best known signal 
model. We have provided critical insight into the role of higher harmonics in the fast gravitational-wave phasing, higher PN corrections in the amplitudes and their interplay with the slow modulations induced by LISA's motion (see Appendix B).

By using a waveform whose phase is correct to order $v^7$ (3.5PN) and
with the amplitudes correct to order $v^5$ (2.5PN), we have found that LISA's
angular resolution improves typically by a factor of 10, even with a very conservative choice for the detector's lower cutoff frequency (i.e., $10^{-4}$ Hz). This means
that LISA will be able to uniquely identify the galaxy cluster in which
the merger event took place and thereby facilitate optical identification
and the measurement of the redshift of the source. Together with the fact
that binary SMBH sources are standard sirens, this means that LISA will
be able to measure the cosmological parameters by circumventing the
lower rungs of the cosmic distance ladder. 

\begin{acknowledgments}
For useful discussions and encouragement we would like to thank 
Karsten Danzmann and Bernard Schutz. We benefited from discussions
on the importance of weak lensing in the context of LISA with
Daniel Eisenstein, Daniel Holz and Scott Hughes. We are grateful to
Miquel Trias and Alicia Sintes for sharing with us their results,
which corroborate ours.
B.R. Iyer thanks Cardiff University and IHES, France for
hospitality during different stages of this work. This research was supported in part by PPARC grant PP/B500731/1. The calculations leading to the results of this paper were performed with the aid of {\tt Mathematica}.
\end{acknowledgments}

\begin{appendix}

\section{Implementation of the parameter estimation with FWF and issues related to numerical accuracy}

The computation of the covariance matrix for LISA
using FWF starts with a simpler code (in {\tt Mathematica})
which reproduces the results of Ref.~\cite{Chris06,ChrisAnand06,ChrisAnand06b}
for the ground-based detectors.
The modulations due to LISA's orbital motion are then accounted for 
to obtain the waveform in LISA's frame. This waveform 
is differentiated analytically and the derivatives are stored in an array and used
in the computation of the various elements of the Fisher matrix.
The  integrations that are needed in computing different elements of
the Fisher information matrix are performed using numerical 
integration routines of {\tt Mathematica}.
The program thus evaluates the Fisher matrix for the given input
values of the component masses of the binary, the four angles
describing the source in the solar system barycenter frame
and the PN order of the amplitude for a fixed distance of 3 Gpc 
(phasing is always 3.5PN).
The input masses are always physical masses consistent
with the cosmological model we described earlier.

We have checked that our code reproduces the numbers in Ref.~\cite{Cutler98} 
(with the corresponding restricted waveform which uses a 1.5PN phasing including the
 spin-orbit parameters, and the corresponding noise
PSD) with less than $1\%$ difference. Since we use the
noise PSD of Ref.~\cite{BBW05a}, we have checked that we
obtain the numbers for the pattern-averaged case of \cite{BBW05a}
to less than 1\% difference using their waveform
parametrization which includes all nonspinning and spinning binary
parameters up to 2PN but ignores
higher PN order terms in the phasing. Lastly, we recover the results
of \cite{ALISA06}, which uses a 3.5PN phasing for the parameter estimation
and recover the results again to less than 1\% difference.

The numbers quoted in this paper were checked with three independent
codes, all in {\tt Mathematica}, but using different numerical
integration routines (such as {\tt NIntegrate}
and {\tt ListIntegrate}) and matrix inversion routines (such
as {\tt Inverse} and another method based on singular value decomposition 
(SVD)), and all three codes agree to less than $5\%$.

A large dimension of the parameter space often leads to {\it ill-conditioned }
Fisher matrices, i.e., the magnitude of the ratio
of the smallest and the largest of its singular values (the inverse of the
{\it condition number}) approaches the machine's floating point precision
($10^{-16}$ in our case). 
For the ($6.45 \times 10{^4},1.29 \times 10{^6})M_\odot$ system, only the RWF
Fisher matrix is ill-conditioned, for the third angle. The
($10{^5}, 10{^6})M_\odot$ system, for both RWF and FWF, is free of
ill-conditioned Fisher matrices. More specifically, the inverse of the
condition number is always $10^{3}-10^{5}$ times larger than $10^{-16}$.
The {\tt Mathematica} inversion routine as well as singular value
decomposition (SVD) are used to obtain the covariance matrix and both
methods give the same results.
For these systems, we also note that the numerical
eigenvalues (computed using the {\tt Mathematica} function {\tt Eigenvalues}) coincide with the numerical singular values (obtained through {\tt SingularValueDecomposition})
at the standard floating point precision of $10^{-16}$.
On the other hand, the Fisher matrices for the $(10{^6}, 10{^7})M_\odot$ system show ill-conditioned behaviour at this precision. However, this may not mean that the results obtained are unreliable for reasons next outlined. The SVD of the Fisher matrices shows that one or more of the singular
values approach zero. Indeed, a machine precision calculation of the singular values yields zero for at least one singular value of the Fisher matrices for this system. In such cases, we replace the singular values by the numerical eigenvalues (whose machine precision calculation does not yield zero) to obtain the inverse using SVD.
However, {\tt Mathematica } has the option of computing at higher precision by padding unknown digits beyond the known ones. Using this facility and
repeating the inversion procedure using SVD but with precision higher than $10^{-16}$ does {\it not} alter the results and shows that the singular values are equal to the eigenvalues, as expected,
with none of them being zero. The condition number is also seen to be 10 times larger than the precision used. In the case of one or more almost zero singular values (at the standard floating point precision), one can also obtain a pseudoinverse \cite{FortranRecipe} that is closest to the ``real" inverse, in the least-squares sense, and end up with different results. However, for the $(10^{6},10^{7})M_\odot$ system, the errors given by the pseudoinverse do not seem physically correct
and so we refrain from using it in this work. A detailed analysis, by perturbing the Fisher
matrices and observing how the inverses behave will be a stronger test of the reliability of our results
\cite{Vallisneri07}.

A recent Markov Chain Monte Carlo (MCMC) analysis of
 Ref.~\cite{CornishPorter06} revealed an interesting point relevant
to our analysis. It compared the estimates from the Fisher information matrix
with those obtained from MCMC and found that there is excellent
agreement between the two methods in the case of the extrinsic parameters 
(what we refer to as ${\mathbf p}_{\rm slow}$), although in the case of the 
intrinsic parameters the two methods are not in good agreement. This is
good news since our main concern here is the angular resolution and the
luminosity distance. Though their analysis used the restricted waveform, 
we believe that similar trends would exist in the case of FWF too
since the dimensionality of the parameter space is the same as that of RWF.

\section{Parameter estimation with the FWF and variation with PN order in amplitude}

In this Appendix we will attempt to provide some qualitative understanding of the trends 
observed in the parameter estimation accuracy with increasing PN order in amplitude.
Recall that the covariance matrix goes as 
$1/\rm SNR^2$ and therefore errors go down as $1/\rm SNR.$ However, 
notice from Table \ref{tab:PE1} that even though the FWF SNR is less than the RWF SNR all errors 
decrease as we go from RWF to FWF. This provides the clue
that the SNR cannot account for the improved performance of FWF: In going from 
the  RWF to 0.5PN, the ratios of the two SNRs are nowhere close to the inverse 
ratios of the errors in various parameters. Further, as we go from 0.5PN 
to 2.5PN, although the products of the SNR with errors do not fluctuate much, 
they are  noticeably different implying that the SNR is not the main reason 
why the parameter estimation improves. 

Next, we note that at 0.5PN both the first and the third harmonic 
possess the same structure ($f$ dependence) and almost the same variety
(dependence on the inclination angle $\iota$) (compare Eq.~(5.7b) and (5.8b) of \cite{ABIQ04}).
However, their effects on PE, when examined independently, are vastly
different as seen in  Table \ref{tab:PE2}.
\begin {table}[t]
\centering
\begin{tabular}{|l|c|c|c|c|c|c|c|c|}
\hline
\hline
 PN  & \multicolumn {8}{c|}{$(10^6,\, 10^7) M_\odot$; {\bf $z=0.55$ }; 
 $\mu_{\rm S}=-0.8,\phi_{\rm S}=1,\mu_{\rm L}=0.5,\phi_{\rm L}=3$ (A3).}\\
\cline{2-9}
order& \multicolumn{1}{c|}{SNR} & \multicolumn {1}{c|}{~$\Delta\,\ln{\cal M}$~} & \multicolumn {1}{c|}{~~~$\Delta\,\delta$~~~} & \multicolumn {1}{c|}{~~~$\Delta\,t_{\rm C}$~~} & \multicolumn {1}{c|}{~~$\Delta\,\phi_{\rm C}$~~~} & \multicolumn{1}{c|}{~$\Delta \ln D_{\rm L} $~} & \multicolumn {1}{c|}{$\Delta\,\Omega_{\rm S}$} & \multicolumn {1}{c|}{$\Delta\,\Omega_{\rm L}$}\\
 & & $(10^{-6})$ &  $(10^{-6})$ & (sec) & (rad) & ($10^{-3}$) & ($10^{-5}$ str) & ($10^{-5}$str)\\
\hline
0                         & 1824 & 380 & 310 & 90 & 65  & 32  & 2400 & 6600 \\
0.5                       & 2005 & 110 & 110 & 32 & 2.6 & 2.3 & 6.3  & 9.9\\
0.5 (3rd)                  & 2004 & 110 & 110 & 32 & 2.6 & 2.3 & 6.3  & 10 \\
0.5 (1st)                  & 1825 & 370 & 310 & 81 & 21  & 10  & 360  & 960\\
0.5 (2nd)                  & 1648 & 410 & 340 & 90 & 12  & 10  & 690  & 1800\\
0.5 (2nd with span of 3rd) & 1757 & 190 & 140 & 31 & 5.0 & 2.2 & 17   & 44\\
0.5 (7th)                  & 1943 & 11  & 22  &8.5 & 1.9 & 2.1 & 3.6  & 4.2\\
\hline
\hline
\end{tabular}
\caption{
Parameter estimation at 0.5PN order  with mock waveforms corresponding to
different choices. The first (second) row corresponds to the restricted 
(0.5PN) results. The third (fourth) row corresponds to the 0.5PN waveform 
with only the 3rd (1st) harmonic and suppressing the 1st (3rd) harmonic. 
The fifth row  corresponds to a mock waveform suppressing the 
first harmonic and replacing the third harmonic by the second.
This implies retaining its  polarisation amplitude and phase 
but changing $t(f/3)\rightarrow t(f/2)$ in the Fourier and 
the Doppler phases and decreasing the frequency span 
from $3(F_{\rm LSO}-F{\rm in})$ to $2(F_{\rm LSO}-F_{\rm in})$.
The sixth row corresponds to the previous case but with
an enhanced span  $3(F_{\rm LSO}-F_{\rm in})$.
The last row corresponds to a mock waveform
where  the seventh harmonic is moved from 2.5PN to 0.5PN 
and treated with its associated normal frequency span $7(F_{\rm LSO}-
F_{\rm in})$ with the 1st and 3rd harmonics suppressed. In this and other
tables in this appendix the system and angles are the same as in Table \ref{tab:PE1}.
}
	\label{tab:PE2}
\end {table}
   The 0.5PN PE is hardly distinguishable from the PE with the  first harmonic
suppressed and thus the improvement at 0.5PN  is solely due to the third harmonic.
This is further confirmed by performing the PE with a ``mock" FWF where
the third harmonic is replaced by the second harmonic: as seen from 
Table \ref{tab:PE2}
the PE worsens in this case. Thus, the presence of higher harmonics
brings about improved PE. It must, however, be pointed out that 
the improvement is obtained {\it not} merely due to the presence of the
higher harmonic but {\it crucially} due to the {\it increased
span} of the $k$th harmonic in the frequency domain ($k(F_{\rm LSO} - F_{\rm in})$).
This is first  checked by working with a mock 0.5PN waveform containing the 
second harmonic but using for its span the increased span of the third.
It is confirmed by an alternative mock waveform, 
suppressing the first and third harmonics and including instead
 the seventh harmonic at 0.5PN. The dramatic improvement in PE compared to the
regular 0.5PN waveform, but with almost the same SNR, shows that at the 0.5PN
level, where complications due to PN corrections to the harmonics and higher
order expansion coefficients of $\dot{F}^{-1/2}$ are not present,
 the SNR is not the determining factor.

      To understand the  variation in PE  from 0.5PN to 2.5PN
we need to disentangle many effects. The FWF introduces more
structure ($f$ dependences) at different PN orders.
Different harmonics appear at different PN orders and
in our model are associated with their respective spans.
At 2.5PN, higher order corrections appear in every harmonic except
the sixth and the seventh harmonics. The frequency sweep $\dot{F}^{-1/2}$ 
in the Fourier transform of the binary signal brings in further
PN corrections associated  with it (the coefficients $S_{(m/2)}$ in Eq.~(\ref{FWF})). 
Finally, the coefficients 
involve angles which are fixed in the barycentric
frame but suffer modulations (Doppler and orientational) due to LISA's motion. 
To this end, we reexamined the PE with a series of mock waveforms
including where possible only one of the above aspects.
For instance, in Table  \ref{tab:PE3} we consider only the second harmonic
and its associated PN corrections.
\begin {table}[h]
\centering
\begin{tabular}{|l|c|c|c|c|c|c|c|c|}
\hline
\hline
 PN  & \multicolumn {8}{c|}{$(10^6,\, 10^7) M_\odot$; {\bf $z=0.55$ };
 $\mu_{\rm S}=-0.8,\phi_{\rm S}=1,\mu_{\rm L}=0.5,\phi_{\rm L}=3$ (A3).}\\
\cline{2-9}
order& \multicolumn{1}{c|}{SNR} & \multicolumn {1}{c|}{~$\Delta\,\ln{\cal M}$~} & \multicolumn {1}{c|}{~~~$\Delta\,\delta$~~~} & \multicolumn {1}{c|}{~~~$\Delta\,t_{\rm C}$~~} & \multicolumn {1}{c|}{~~$\Delta\,\phi_{\rm C}$~~~} & \multicolumn{1}{c|}{~$\Delta \ln D_{\rm L~}$~} & \multicolumn {1}{c|}{$\Delta\,\Omega_{\rm S}$} & \multicolumn {1}{c|}{$\Delta\,\Omega_{\rm L}$}\\
 & & $(10^{-6})$ &  $(10^{-6})$ & (sec) &(rad) & ($10^{-3}$) & ($10^{-5}$ str) & ($10^{-5}$ str)\\
\hline
0, 0.5   & 1824 & 380 & 310 & 90  & 65  & 32 & 2400  & 6600 \\
1, 1.5   & 1510 & 450 & 370 & 110 & 110 & 44 & 2000  & 5900 \\
2        & 1387 & 490 & 400 & 120 & 120 & 45 & 2300  & 6700 \\
2.5      & 1353 & 500 & 410 & 120 & 130 & 48 & 2400  & 7200 \\
\hline
\hline
\end{tabular}
\caption{Variation of PE with PN orders in amplitude for a mock FWF
retaining only the 2nd harmonic and its higher order PN corrections. 
The PN corrections to a given harmonic at PN order $n/2$ add terms
of type $(2\pi M f/k)^{n/3}$. While they bring in new structure 
in the waveform they do not help improve PE; instead they 
enhance the covariances among different parameters and thereby 
worsen PE relative to RWF.
}
	\label{tab:PE3}
\end {table}
From the results it is clear that although PN corrections to the 2nd
harmonic bring in additional terms, they do not improve PE; rather, 
in comparison with the RWF, the PE worsens. This is probably due to the
fact that PN corrections with the same frequency dependence  as before
(or powers thereof) increase the covariances among the various 
parameters and thus worsen PE.

In Table \ref{tab:PE5}, on the other hand, the mock waveform includes
all the higher harmonics but excludes their PN amplitude corrections and PN corrections arising from the frequency sweep. Clearly, the PE
improves monotonically as we go from one order to the next in all the
parameters quoted and so does the SNR.
\begin {table}[ht]
\centering
\begin{tabular}{|l|c|c|c|c|c|c|c|c|}
\hline
\hline
 PN  & \multicolumn {8}{c|}{$(10^6,\, 10^7) M_\odot$; {\bf $z=0.55$ };~$\mu_{\rm S}=-0.8,\phi_{\rm S}=1,\mu_{\rm L}=0.5,\phi_{\rm L}=3$ (A3).}\\
\cline{2-9}
order& \multicolumn{1}{c|}{SNR} & \multicolumn {1}{c|}{~$\Delta\,\ln{\cal M}$~} & \multicolumn {1}{c|}{~~~$\Delta\,\delta$~~~} & \multicolumn {1}{c|}{~~~$\Delta\,t_{\rm C}$~~} & \multicolumn {1}{c|}{~~$\Delta\,\phi_{\rm C}$~~~} & \multicolumn{1}{c|}{~$\Delta \ln D_{\rm L} $~} & \multicolumn {1}{c|}{$\Delta\,\Omega_{\rm S}$} & \multicolumn {1}{c|}{$\Delta\,\Omega_{\rm L}$}\\
 & & $(10^{-6})$ &  $(10^{-6})$ & (sec) &(rad) & ($10^{-3}$) & ($10^{-5}$str) & ($10^{-5}$str)\\
\hline
0   & 1824 & 380   & 310 & 90 & 65  & 32 & 2400 & 6600 \\
0.5 & 2005 & 110   & 110 & 32 & 2.6 & 2.3 & 6.3 & 9.9\\
1   & 2058 &  83   & 87  & 26 & 2.4 & 2.2 & 5.1 & 7.03 \\
1.5 & 2068 &  78   & 82  & 25 & 2.3 & 2.2 & 4.8 & 6.4\\
2   & 2070 &  77   & 82  & 24 & 2.3 & 2.2 & 4.8 & 6.3\\
2.5 & 2070 &  77   & 82  & 24 & 2.3 & 2.2 & 4.8 & 6.3\\
\hline
\hline
\end{tabular}
\caption{Variation of PE with PN orders  for a mock FWF
 retaining  all  the harmonics 
and neglecting  both   higher order PN  corrections to them 
 and also  PN corrections to  $(\frac{dF}{dt})^{-1/2}$ arising
from the frequency sweep.
Thus most  higher order PN corrections to the harmonics are neglected.
Higher harmonics generally improve parameter estimation. 
PN corrections to harmonics tend to degrade PE.
}
\label{tab:PE5}
\end {table}
Additionally, an examination of the results in Tables \ref{tab:PE1}
and \ref{tab:PE5}  reveals that while the higher harmonics by 
themselves tend to generically improve PE, the PN 
corrections to the harmonics arising from the higher PN amplitudes
and higher PN frequency sweep tend to degrade PE.
However, their effect is less dramatic than one may naively expect.
This is because higher harmonics necessarily appear at higher
PN orders and, as is evident from Eq.~(\ref{FWF}), they come with higher
powers of the PN expansion parameter: $(2\pi M f_k)^{n/3}$ for the $k$th harmonic appearing at the $n/2$th PN order. The upper cutoff
of the $k$th harmonic in the frequency domain is $k\,F_{\rm LSO}$ at which 
$(2\pi M f)^{1/3}$ reaches its maximum of $6^{-1/2}$.
The 7th harmonic, for example, will be scaled by a
factor which is always less than $6^{-5/2}$ and
 consequently will contribute less power.
The seventh harmonic in the FWF at its regular PN order
hardly leads to any improvement in PE. However, as remarked earlier
in Table \ref{tab:PE2},  had it appeared at 0.5PN its impact would have
been substantial. We have verified these trends for another angle  and for 
a lower mass system of $(10^5,\, 10^6)M_\odot$; for the sake of
brevity we leave out the details.

\begin {table}[ht]
\centering
\begin{tabular}{|l|c|c|c|c|c|c|c|c|c|}
\hline
\hline
\multicolumn {10}{|c|}{$(10^6,\, 10^7) M_\odot$;~Correlation-Coefficient Matrix;~$0$PN (A3)}\\
\cline{1-10}
& \multicolumn{1}{c|}{~$\ln{\cal M}$~} & \multicolumn{1}{c|}{~$\delta $~} & \multicolumn {1}{c|}{~$t_{\rm C}$~} & \multicolumn {1}{c|}{~$\phi_{\rm C}$~~} &\multicolumn{1}{|c|}{~$ \ln D_{\rm L} $~} & \multicolumn {1}{c|}{~$\mu_{\rm S}$~} & \multicolumn {1}{c|}{~$\phi_{\rm S}$~} & \multicolumn {1}{c|}{$\mu_{\rm L}$} & \multicolumn {1}{c|}{$\phi_{\rm L}$}\\
       
\hline
$\ln{\cal M}$ & 1    &$-0.99$ & 0.93 &$-0.19$ &$-0.17$ & 0.14  &$-0.13$ &$-0.15$ & 0.031 \\
\hline
$\delta $     &      &  1     &$-0.92$ & 0.098  & 0.084 &$-0.06$ & 0.065  & 0.074 &$-0.0088$ \\
\hline
$t_{\rm C}$   &      &        &  1     &$-0.43$ &$-0.45$ & 0.33 &$-0.43$ &$-0.46$ &$-0.022$ \\
\hline
$\phi_{\rm C}$ &      &        &        &   1    & 0.98 &$-0.95$ & 0.70 & 0.92 &$-0.39$ \\
\hline
$\ln D_{\rm L}$&     &        &         &        &    1   &$-0.88$ & 0.81 & 0.98  &$-0.22$\\
\hline
$\mu_{\rm S}$ &     &         &        &        &        &    1   &$-0.45$ &$-0.77$ & 0.65 \\
\hline
$\phi_{\rm S}$&     &         &        &        &        &        &   1    & 0.92 & 0.38 \\
\hline
$\mu_{\rm L}$ &     &         &        &        &        &        &        &    1   &$-0.016$ \\
\hline
$\phi_{\rm L}$&     &         &        &        &        &        &        &        &    1   \\
\hline
\hline
\end{tabular}
\caption{Correlation coefficients computed using RWF for a binary at $z\simeq 0.55$
comprising $(10^6,\, 10^7) M_\odot$ SMBH.  The Table shows how the 
`fast' and `slow' variables behave.
There is a high correlation  among parameters in the same subclass,
but only a weak correlation between members of different subclasses. 
Entries that are vacant can be found by symmetry.  
}
\label {tab:PE6}
\end {table}
\begin {table}[ht]
\centering
\begin{tabular}{|l|c|c|c|c|c|c|c|c|c|}
\hline
\hline
\multicolumn {10}{|c|}{$(10^6,\, 10^7) M_\odot$;~Correlation-Coefficient
Matrix;~$2.5$PN (A3)}\\
\cline{1-10}
& \multicolumn{1}{c|}{~$\ln{\cal M}$~} & \multicolumn{1}{c|}{~$\delta $~} &
\multicolumn {1}{c|}{~$t_{\rm C}$~} & \multicolumn {1}{c|}{~$\phi_{\rm C}$~~}
&\multicolumn{1}{|c|}{~$ \ln D_{\rm L} $~} & \multicolumn {1}{c|}{~$\mu_{\rm
S}$~} & \multicolumn {1}{c|}{~$\phi_{\rm S}$~} & \multicolumn {1}{c|}{$\mu_{\rm
L}$} & \multicolumn {1}{c|}{$\phi_{\rm L}$}\\

\hline
$\ln{\cal M}$ &  1  &$-0.97$ & 0.94 & 0.37 &$-0.075$  &$-0.0069$ &$-0.10$ &$-0.12$ &$-0.097$ \\
\hline
$\delta $     &     &    1    &$-0.99$ &$-0.54$ & 0.06 & 0.030 & 0.093 & 0.093 & 0.11 \\
\hline
$t_{\rm C}$   &     &         &    1   & 0.63 &$-0.10$ &$-0.016$ &$-0.14$ &$-0.13$ &$-0.14$ \\
\hline
$\phi_{\rm C}$ &     &         &        &     1  & 0.051 &$-0.17$ & 0.015 & 0.16 &$-0.18$  \\
\hline
$\ln D_{\rm L}$&     &         &        &        &    1   &$-0.13$ & 0.90 & 0.71  & 0.47\\
\hline
$\mu_{\rm S}$ &     &         &        &        &        &    1   & 0.036 &$-0.56$ & 0.68 \\
\hline
$\phi_{\rm S}$&     &         &        &        &        &        &   1 & 0.76 & 0.71 \\
\hline
$\mu_{\rm L}$ &     &         &        &        &        &        & &    1 & 0.18  \\
\hline
$\phi_{\rm L}$&     &         &        &        &        &        & & &    1   \\
\hline
\hline
\end{tabular}
\caption{Correlation coefficients computed using FWF for a binary at $z\simeq 0.55$,
comprising $(10^6,\, 10^7) M_\odot$ SMBH.
The Table  clearly shows a more pronounced decoupling between the 
 `fast' and `slow' variables at 2.5PN.
 There is a high correlation among elements of the same subclass but a  
 lower correlation between the fast and slow subclasses.}
\label {tab:PE7}
\end {table}

Our  last comment relates to  the decoupling between the fast variables
${\mathbf p}_{\rm fast}$ and the slow variables  ${\mathbf p}_{\rm slow}$ 
in the case of the FWF. We have  checked that in the case of the FWF 
the correlation coefficients are very small (of the order of $0.1$-$0.3$)
between these two subsets of parameters (see Table \ref{tab:PE6} and Table \ref{tab:PE7}).
There exists, however, a high correlation among different parameters of the same
type as in the case of RWF (see, e.g., Ref.\ \cite{Hughes02}).
Specifically, as in the RWF case ~\cite{Hughes02}, the distance estimation 
could be improved if the source is better localized in the sky.

It is worth recalling that in the parameter estimation problem we are not only limited by statistical errors due to noise but also by theoretical or systematic errors arising at \emph{any} PN order due to the limited accuracy of the waveforms. Recently Cutler and Vallisneri \cite{CutlerVallisneri07} have looked into this issue more critically. In the context of the present work, we would like to stress that despite the fact that the largest improvement in PE arises from the third harmonic at 0.5PN, the need to limit systematic errors mandates the use of the best available waveform, i.e., the one at 2.5PN in amplitude and 3.5PN in phase.

\end{appendix}

\end{document}